\documentstyle[preprint,aps,prl]{revtex}

\begin{document}
\draft
\title{Electrically Tunable Collective Modes in a MEMS Resonator Array}
\author{E. Buks and M. L. Roukes}
\address{Condensed Matter Physics 114-36, California Institute of Technology,
Pasadena, CA 91125}
\date{\today}
\maketitle

\begin{abstract}
Using optical diffraction, we study the mechanical vibrations of an array of
micromechanical resonators. \ Implementing tunable electrostatic coupling
between the suspended, doubly-clamped Au beams leads to the formation of a
band of collective vibrational modes within these devices. \ The evolution
of these modes with coupling strength is clearly manifested in the optical
diffraction pattern of light transmitted through the array. \ The
experimental results are analyzed using a simple model for one-dimensional
phonons. \ These structures offer unique prospects for spectral analysis of
complex mechanical stimuli.
\end{abstract}


\hsize\textwidth\columnwidth\hsize\csname@twocolumnfalse\endcsname

Interaction between light and mechanically vibrating systems was first
discussed by Einstein and Bohr. \ A gedanken experiment of optical
diffraction by a vibrating two slit structure was employed to demonstrate
the complementarity principle of quantum mechanics \cite{200}. \ These ideas
were later elaborated in order to formulate the theory of x-rays and
neutrons diffraction by crystals \cite{qtss} \cite{ssp}. \ In these systems
mechanical vibrations due to both thermal and quantum fluctuations strongly
modify the diffraction pattern. \ This allows experimental study of the
crystal's normal modes of vibration (phonons) and their dispersion relation.
\ Micro electro-mechanical systems (MEMS) technology allows studying similar
phenomena using artificial mechanical systems at mesoscopic length scales. \
Such studies are motivated not only by scientific interest but rather also
by the prospects of developing new micro opto-mechanical devices. \ This
rapidly growing field of research employs MEMS technology to realize a
variety of on-chip fully integrated optical devices (for a review see \cite
{mod}). \ A particular example related to the present work is the movable
diffraction grating. \ In this device the efficiency of diffraction is
controlled by moving the grating with respect to the substrate beneath it,
allowing thus optical modulation \cite{688} \cite{145} \cite{1165}. \
Operation of such devices at relatively high frequencies however requires a
good understanding of the dynamics and mechanical properties of the system.

In the present paper we use optical diffraction to study the mechanical
properties of a periodic array (grating) of suspended doubly-clamped beams
made of Au. \ What is novel and especially interesting about the present
work is that our devices allow application of mutual electrostatic forces 
{\it between} the beams. This coupling \cite{512} gives rise to the
formation of a band of collective modes of vibration (phonons). \ We excite
these collective modes parametrically and employ optical diffraction to
study the response. \ A simple model describing our system is presented and
compared with experiment. \ We conclude by briefly describing some unique
device applications of this system.

The bulk micromachining process employed for sample fabrication is described
in Fig. 1. \ In this process the substrate beneath the grating is completely
etched away, thus allowing optical access to the grating from both sides. \
In the first step chemical vapor deposition is employed to deposit a 70 $%
\mathop{\rm nm}%
$ thick layer of low-stress silicon nitride on the front and back sides of a
Si wafer. \ A square window is opened in the silicon nitride on the back
side using photolithography and wet etching (Fig. 1(a)). \ The structure
shown in Fig. 1(b) is realized through a highly selective, anisotropic KOH
etch for the backside of the Si wafer. This occurs within the patterned
region and yields a 270$\mu $m square silicon nitride membrane on the front
side of the wafer. \ The grating beams and adjacent electrodes are
fabricated on top of this membrane using electron beam lithography, followed
by thermal evaporation of Au and liftoff (Fig. 1(c)). \ Each resulting beam
has length $b=270$ $\mu $m, width $1\mu $m and thickness $0.25\mu $m
(measured using an atomic force microscope) and the grating period is $%
a=4\mu $m. \ In the final step the membrane is removed using electron
cyclotron resonance (ECR) plasma etching from the back side of the sample. \
This process step employs an Ar/NF$_{3}$ gas mixture, and results in
suspension of the Au beam array (Fig. 1(d)). \ Figure 1(e) shows a side view
micrograph of the device. \ The electrodes form two interdigitated combs;
with fingers alternately connected to the two base electrodes. \ This design
allows application of electrostatic forces between the beams.

To characterize uniformity within the device we measure the fundamental
resonance frequency of {\it each} suspended beam in the array. \ This is
done {\it in-situ,} using the output from a commercial scanning electron
microscope's imaging system to detect mechanical displacement. \ We have
employed this technique previously to study the mechanical properties of
individual, similarly-fabricated Au beams \cite{p1} \cite{p2}. Fig. 2(a)
shows a typical response peak from an individual beam. \ By measuring all 67
suspended beams in the array we find that the distribution of resonance
frequencies has an average of 179.3 kHz and a standard deviation of 0.53 kHz
(see Fig. 2(b)). \ Individual mechanical quality factors $Q$ range from
2,000 to 10,000. \ Note that no correlation is found between the location of
the beam within the device and its specific resonance frequency or $Q$; the
small beam-to-beam variations appear to be random.

What is expected when a voltage $V$ is applied between the two combs? \ We
employ a simple one-dimensional model for an $N$-element array of coupled
pendulums \cite{ssp} to describe our system (see Fig. 3(a)). \ While the
first and last pendulums in the array are clamped and stationary, all others
($n=2,3,...,N-1$) are free to oscillate about their equilibrium positions $%
na $. \ Here $a$ represents the equilibrium spacing between neighboring
pendulums. \ In the absence of any coupling, the angular frequency for small
oscillations of each (identical) pendulum is $\omega _{0}$. \ The
displacement of the system is described by a set of coordinates $x_{n}$ ($%
n=1,2,...,N$) (see Fig. 3(a)). \ Applying a voltage $V$ gives rise to an
attractive interaction between each pendulum and its nearest neighbors $%
\varphi \left( s\right) =-C\left( s\right) V^{2}/2$, where $s$ is the
distance between the interacting pendulums and $C$ is the capacitance. \
Neglecting coupling between non-neighboring pendulums, and assuming small
oscillations, we find the following set of equations of motion:

\begin{equation}
m\stackrel{\cdot \cdot }{x}_{n}=-m\omega _{0}^{2}x_{n}+u\left(
2x_{n}-x_{n-1}-x_{n+1}\right) ,  \label{eom}
\end{equation}
where $n=2,3,...,N-1$, and $u=-\varphi ^{^{\prime \prime }}\left( a\right) $
(here dots represents time derivatives and primes represent spatial
derivatives). \ Note that $u>0$ due to the attractive nature of the
interaction between nearest neighbors. \ These equations can be greatly
simplified by employing a transformation to the eigenmodes (phonons) of the
system:

\begin{mathletters}
\begin{equation}
x_{n}=\sum_{m=2}^{N-1}\chi _{n}^{(m)}v_{m},  \label{tra}
\end{equation}
where $\chi _{n}^{(m)}=\sqrt{2/\left( N-1\right) }\sin \left( k_{m}\left(
n-1\right) a\right) $ is the spatial shape of mode number $m$ ($m=2,3,...,N-1
$), $\ k_{m}=\left( m-1\right) \pi /L$ is the wavevector, and $L=\left(
N-1\right) a$ is the length of the system. \ Figure 3(b) shows the shape of
the three lowest modes $m=2,3,4$. \ Substituting Eq. (\ref{tra}) in Eq. (\ref
{eom}) leads to a set of {\it decoupled} equations of motion:

\end{mathletters}
\begin{equation}
\stackrel{\cdot \cdot }{v}_{m}=-\omega _{m}^{2}v_{m},  \label{uncoup}
\end{equation}
where $\omega _{m}^{2}\left( V\right) =\omega _{0}^{2}-\left( 2C^{^{\prime
\prime }}\left( a\right) /m\right) V^{2}\sin ^{2}\left( k_{m}a/2\right) $. \
A stationary voltage $V=V_{dc}$ thus gives rise to the formation of a band
of collective modes between frequencies $\omega _{0}$ and $\omega _{b}=\sqrt{%
\omega _{0}^{2}-\left( 2C^{^{\prime \prime }}\left( a\right) /m\right)
V_{dc}^{2}}$. \ The associated wavevectors, $k_{m}$, vary from zero to $\pi
/a$ (see Figure 3(c)).

Each mode can be selectively excited by adding an AC voltage to the DC bias,
namely $V=V_{dc}+V_{ac}\cos \left( \gamma t\right) $. \ Assuming $%
V_{ac}<<V_{dc}$ we find from Eq. (\ref{uncoup}):

\begin{equation}
\stackrel{\cdot \cdot }{v}_{m}=-\omega _{m}^{2}\left( V_{dc}\right) \left(
1-h_{m}\cos \left( \gamma t\right) \right) v_{m},  \label{par}
\end{equation}
where $h_{m}=2\left( V_{ac}/V_{dc}\right) \left[ \left( \omega _{0}/\omega
_{m}\right) ^{2}-1\right] $. \ Thus an AC voltage component gives rise to
parametric excitation of each mode with amplitude $h_{m}$ \cite{mech} \cite
{nlo} \cite{math}. \ Parametric resonance occurs when the frequency of the
AC voltage $\gamma $ is close to $2\omega _{m}/l$, where $l$ is an integer.
\ Near these values the system may exhibit unstable behavior in which the
amplitude of oscillations grows as a function of time. \ In the linear
theory of parametric resonance this growth is exponential, and occurs when
the amplitude of parametric excitation $h$ exceeds a critical value
dependent upon the damping in the system. \ Most systems, however, possess
some degree of non-linearity which comes into play as soon as the amplitude
of the motion becomes appreciable. \ Thus, while the linear theory is useful
in determining the conditions for the occurrence of parametric resonance, it
is inadequate for determining the steady-state response of the system. \
Unfortunately, the nonlinear coefficients of our system are not known and
therefore its steady-state response cannot be predicted. \ However, we
expect that the response of modes with high index $m$ will be relatively
large because $h_{m}$ increases in magnitude with $m$.

We detect the collective mechanical vibrations of the array by diffraction
measurements. \ With spatially-uniform light incident upon the array, the
intensity of diffraction is proportional to the form factor $\left| \phi
\left( q_{x}\right) \right| ^{2}$ where:

\begin{equation}
\phi \left( q_{x}\right) =\sum_{n=1}^{N}\exp \left[ iq_{x}\left(
na+x_{n}\left( t\right) \right) \right] .  \label{phi}
\end{equation}
Here $q_{x}$ is the $x$ component of the change in wavevector between
incoming and outgoing waves (the $x$ direction lies in the plane of the
sample and is perpendicular to the long axis of the beams) \cite{four}. \
Consider the case where the system is tuned to a diffraction peak, namely $%
q_{x}a=2\pi l$ with $l$ an integer. \ As shown below, for this particular
case the interpretation of the experimental results becomes greatly
simplified. \ We calculate the form factor for the case where mode $m$
oscillates with amplitude $A_{m}$ and all other modes are stationary. \
Assuming small oscillations, namely $q_{x}x<<1$, we find:

\begin{equation}
\left| \phi \left( q_{x}\right) \right| ^{2}=N^{2}+\left[ q_{x}A_{m}\cot
\left( k_{m}a/2\right) \cos \left( \omega _{m}t\right) \right] ^{2}\frac{%
\left( -1\right) ^{m}+1}{N-1}.  \label{phi m}
\end{equation}
Thus the collective mechanical oscillations give rise to an oscillatory
component in the diffraction signal at angular frequency $2\omega _{m}$. \
Note that with uniform illumination any mode with $m$ odd will not
contribute to diffraction. \ This occurs due to the exact cancelation of the
response from different portions of the entire device. \ Similar cancelation
gives rise to the $\cot ^{2}\left( k_{m}a/2\right) $ factor in Eq. (\ref{phi
m}).

The optical setup utilized for the diffraction measurements is schematically
depicted in Fig. 4. \ A polarization maintaining single mode fiber
(numerical aperture NA$=0.15$ and core diameter $d=9$ $\mu $m) delivers
infrared light from a laser to the sample. \ We use a tunable wavelength
diode laser operating in the range 1535-1635 nm. \ A spherical lens (focal
length $f=1$ mm) collimates the beam, which illuminates the back side of the
grating at normal incidence with respect to the array plane. \ At the array
plane, located a distance $s_{1}=3$ cm from the collimator, we estimate the
diameter of the beam to be $\ 2f$NA$+s_{1}d/f=570$ $\mu $m. \ For the
results presented, the polarization of the incident light is TE, {\it i.e.}
the electric field is parallel to the grating lines. \ We note, however,
that similar results (not presented here) were obtained with perpendicular
polarization. \ For the present case there are four diffraction peaks with
angles $\theta _{n}=\sin ^{-1}\left( n\lambda /a\right) $ ($n=\pm 1,\pm 2$)
with respect to the normal incidence. Here $\lambda $ is the wavelength of
incident light.. \ Transmitted light is collected by a cylindrical lens and
focused into a second, single-mode fiber. \ The lens enables the intensity
of collected light to be maximized without degrading the spectral
resolution. \ The fiber delivers light collected from the first order ($n=1)$
diffraction peak to a photodiode detector. \ The distance between the sample
and the lens, $s_{2}=1$ cm, was chosen as a compromise between two
conflicting considerations, namely, simultaneously maximizing spectral
resolution and light intensity. \ While the former consideration favors a
large distance, the later favors a small one. \ For the distance chosen, the
spectral resolution is $\delta \lambda =a\delta \theta \cos \theta
_{1}=a\left( d/s_{2}\right) \cos \theta _{1}\simeq 3$ nm. \ All measurements
are done at room temperature in vacuum of $\simeq 10^{-3}$ torr.

We first study diffraction from the array in the absence of any
interelectrode bias voltage. \ The inset of Fig. 4 shows the intensity
detected by the photo diode as a function of $\lambda $. \ The full width at
half maximum (FWHM) of the diffraction peak is estimated using the
Fraunhofer diffraction formula \cite{four} to be $\delta \lambda \simeq
\lambda /N=23$ nm. \ We find good agreement between this estimate and the
measured value. \ No effect is resolved from the thermally-driven mechanical
vibrations of the beams. \ Their effect upon diffraction can be
characterized by multiplying the diffracted intensity by a Debye-Waller
factor, $\exp \left( -2W\right) $ \cite{qtss} \cite{ssp}. In the present
case we estimate that $2W\simeq 10^{-8}$, hence thermal fluctuations are not
expected to affect the diffraction from our micromechanical array
significantly. \ This is confirmed by the experimental data.

In order to excite the modes of vibrations of the system externally, we
apply a voltage $V=V_{dc}+V_{ac}\cos \left( \gamma t\right) $ between either
side of the interdigitated electrode arrays. \ We tune the laser wavelength
to the diffraction peak ($\lambda =1582$ nm) and measure the photodetector
response using a lock-in amplifier operating in $2f$ mode. \ This allows us
to detect the Fourier component of the array response at angular
frequency $2\gamma $. \ Figure 5 is a color map plot showing this second
harmonic response, $R,$ as a function of both $V_{dc}$ and $f=\gamma /2\pi $%
. \ The amplitude of the AC voltage is $V_{ac}=50$ mV for these measurements.

With $V_{dc}=0$ we find a peak in $R$ at $f=179.3$ kHz, associated with the
fundamental frequency of the decoupled beams. \ The FWHM of the peak is 0.6
kHz, close to the standard deviation found in the distribution of the
measured fundamental frequencies. \ This leads us to the conclusion that the
width of the response peak at $V_{dc}=0$ is dominated by inhomogeneous
broadening caused by the non-uniformity of the array.

As we increase $V_{dc}$ we observe a gradual increase in the frequency range
where relatively large response is observed; we associate this with the
formation of a band of collective modes. \ The lower frequency bound of this
range $f_{b}=\omega _{b}/2\pi $ (the bottom of the band) for relatively
small $V_{dc}$ is given theoretically by $f_{b}=\left( \omega _{0}/2\pi
\right) \left( 1-C^{^{\prime \prime }}\left( a\right) V_{dc}^{2}/m\omega
_{0}^{2}\right) $. \ A least squares fit to the measured data (see dashed
line in Fig. 5) yields $C^{^{\prime \prime }}\left( a\right) /m\omega
_{0}^{2}=2.7\times 10^{-4}$ V$^{-2}$. \ For comparison we derive below a
rough order of magnitude estimate of this factor. \ We substitute $m$, which
should represent an effective mass of each beam, by the actual mass and we
use the approximation $C^{^{\prime \prime }}\left( a\right) \simeq
\varepsilon _{0}b/g^{2}$, where $\varepsilon _{0}$ is permittivity of free
space and $g$ is the gap between neighboring beams. \ These crude
approximations yield a value of $1.6\times 10^{-4}$ V$^{-2}$ which is quite
close to the value deduced from the experimental data. \ The upper frequency
bound of the band, on the other hand, shows some discrepancy with theory. \
While the measured value depends on $V_{dc}$, our simple model predicts a
upper value that is fixed. \ We obtained similar behavior of this upper
frequency with orthogonal polarization, and with another sample of similar
design.

The rich and detailed structure of the frequency-dependent response $R$
observed in our experiments is not fully understood. \ Experimentally,
Figure 5 shows that $R$ is oscillatory as a function of frequency, and has a
relatively large magnitude close to the lower limit of the band. \
Theoretically this frequency dependence can be found from Eq. (\ref{phi m})
and the density of states (which is determined by the dispersion relation).
\ This leads to an approximated proportionality relation $R\sim
A_{m}^{2}\cos \left( k_{m}a/2\right) /\sin ^{3}\left( k_{m}a/2\right) $ for
even $m$. \ As discussed above, however, our simple{\it \ linear }theory is
not adequate for predicting the steady-state response $A_{m}$. \ Comparison
with experiment leads to the following conclusions: (a) For small $m$ $A_{m}$
goes to zero faster than $m^{3/2}$. (b) Near the lower bound of the band ($%
k_{m}a\rightarrow \pi $) $A_{m}$ grows faster than $\sqrt{1/\cos \left(
k_{m}a/2\right) }$. (c) The spiky behavior indicates that $A_{m}$ is not a
monotonic function of $m$. \ These intriguing details of the device response
warrant further investigation.

Electrically tunable arrays offer unique prospects for opto-mechanical
signal processing devices such as tunable filters and optical modulators. \
A particularly intriguing example is opto-mechanical spectral analysis of
electrical waveforms. \ Consider an arbitrary electrical signal applied
between the two comb electrodes of our device. \ Its Fourier components
falling within the vibrational band (formed by the DC electrostatic coupling
as described above) will parametrically drive the collective modes of the
array. \ Each of these excited modes will result in a diffracted order with
strength directly proportional to the respective Fourier component. \ Since
each order is diffracted at a characteristic angle, {\it continuous,
real-time }spectral analysis of the applied waveform can be realized using a
photodetector array. \ In principle, by scaling the size of the resonant
beams downward into the realm of NEMS\ (nano electromechanical systems), the
operating frequency of such a device can be extended to very high
frequencies \cite{roukes}.

In conclusion, we have demonstrated the ability to induce and control
collective modes of mechanical vibration within an artificial mesoscopic
lattice. \ Further experimental and theoretical work will elucidate what
appear to be rich dynamics in such systems, and will allow their
optimization for novel {\it micro opto-mechanical }device applications.

The authors are grateful to K. Schwab for his assistance in sample
fabrication. \ This research was supported by DARPA MTO/MEMS under grant
DABT63-98-1-0012. \ E. B. gratefully acknowledges support from the
Rothschild fellowship and the R. A. Millikan postdoctoral fellowship at
Caltech.

\begin{figure}[tbp]
\caption{The device is fabricated using bulk micromachining techniques. \ In
steps (a) and (b) a suspended membrane of silicon nitride is formed. \ A
gold beam is fabricated on top of the membrane (c) and the membrane is
etched, leaving the beam suspended (d). \ A side view micrograph of the
device is seen in (e).}
\label{fig:fig1}
\end{figure}

\begin{figure}[tbp]
\caption{(a) Response peak of an individual beam with a resonance frequency
of 179.28 KHz. \ (b) Normalized resonance frequencies of all beams in the
array $(\protect\omega_0-<\protect\omega_0>)/<\protect\omega_0>$ , where $<%
\protect\omega_0>$ is the average value. }
\label{fig:fig2}
\end{figure}

\begin{figure}[tbp]
\caption{(a) A model of $N$ coupled pendulums. \ (b) The shape of the three
lowest modes. \ (c) \ The dispersion relation between the frequency $\protect%
\omega _{m}$ of each mode and the wave vector $k_{m}$. }
\label{fig:fig3}
\end{figure}

\begin{figure}[tbp]
\caption{The optical setup for measurements of the 1st order diffraction
peak of the grating. \ Optical fibers are employed to deliver light to and
from the vacuum chamber where the sample is mounted. \ The DC voltage $%
V_{dc} $ introduces coupling between the beams and the AC voltage $V_{ac}$
is used to parametrically excite the modes of vibration. \ A lock-in
amplifier is employed to measure the response. \ The inset shows the DC
signal of the photo detector as a function of wavelength (for $%
V_{dc}=V_{ac}=0$). }
\label{fig:fig4}
\end{figure}

\begin{figure}[tbp]
\caption{A color map showing relative signal intensities measured at the
lock-in amplifier operating in a $2f$ mode. \ The dependence upon both the
voltage $V_{dc}$ and the frequency of the AC voltage $f$ are shown. \ The
wavelength is tuned to the diffraction peak, namely $\protect\lambda =1582$
nm. \ The dashed white line shows a fit to the measured lower bound
frequency $f_{b}$ of the band. }
\label{fig:fig5}
\end{figure}

\end{document}